\documentclass[showpacs,amssymb,aps]{revtex4}

\usepackage{amsmath}
\usepackage{amssymb}

\begin{document}

\title{Propagators with the Mandelstam-Leibbrandt Prescription in the
  Light-Cone Gauge}  

\author{Ashok Das$^{a}$ and J. Frenkel$^{b}$}
\affiliation{$^{a}$ Department of Physics and Astronomy,
University of Rochester,
Rochester, New York 14627-0171, USA}
\affiliation{$^{b}$ Instituto de F\'{\i}sica, Universidade de S\~{a}o
Paulo, S\~{a}o Paulo, BRAZIL}

\bigskip
%\date{}

\begin{abstract}

We show that the Feynman 
propagator in the light-cone gauge with the Mandelstam-Leibbrandt
prescription 
has a logarithmic growth for large $\tilde{n}\cdot x$ which is
related to the presence of a residual gauge invariance. Furthermore,
we show that the retarded propagator for the $\tilde{n}\cdot A$
component of the gauge field 
develops a coordinate dependent mass. We argue that this
feature is unphysical and may be eliminated by fixing the residual
gauge degrees of freedom. 

\end{abstract}

\pacs{11.15.-q, 11.10.Ef, 12.38.Lg}

\maketitle

\section{Introduction}

In the study of gauge theories, the light-cone gauge has proven to be
quite intriguing and challenging \cite{leibbrandt1}. We recall that in
a  general axial type gauge (where
$n^{2} \neq 0$ in general and which includes both temporal as well as axial
gauges) 
\begin{equation}
n\cdot A = 0,
\end{equation}
the path integral propagator in the momentum space takes the form
\begin{equation}
D_{\mu\nu}^{\rm (PI)} (n,p) = - \frac{i}{p^{2} +
  i\epsilon}\left[\eta_{\mu\nu} - \frac{n_{\mu}p_{\nu} +
  n_{\nu}p_{\mu}}{(n\cdot p)} + \frac{n^{2}}{(n\cdot p)^{2}}\
  p_{\mu}p_{\nu}\right].\label{pipropagator}
\end{equation}
(In the case of non-Abelian theories, the gauge
potential will correspond to a matrix in the adjoint representation
while the propagator will have an identity matrix multiplying the
expression (\ref{pipropagator}).)
The additional poles of the form $\frac{1}{(n\cdot p)}$ in
(\ref{pipropagator}) require a
prescription for the propagator to be well defined. In a recent
paper \cite{das}, we showed that this arbitrariness in the propagator is a
consequence of a residual gauge symmetry in the generating functional
and fixing this determines the propagator uniquely. The explicit form
of the propagator determined in \cite{das} is well behaved for large
values of 
$n\cdot x$. In contrast, in the light-cone gauge where $n^{2}=0$, we
showed that the prescription dependence persists even after a
residual gauge fixing is used. We traced the origin of this
arbitrariness to the presence of an additional local symmetry in the
generating functional.  

We note that the Mandelstam prescription \cite{mandelstam} (the
Leibbrandt prescription \cite{leibbrandt} is equivalent to that of
Mandelstam, but we use the
Mandelstam prescription explicitly through out our discussions) is
widely used in the light-cone gauge calculations. We recall that this 
prescription corresponds to defining the additional pole as
\begin{equation}
\frac{1}{[(n\cdot p)]} = \lim_{\epsilon\rightarrow 0}\ \frac{1}{n\cdot
  p + i\epsilon \tilde{n}\cdot p}.\label{mandelstam}
\end{equation}
Here $\tilde{n}^{\mu}$ is a second light-like vector such that $n\cdot
\tilde{n}\neq 0$.
Such a prescription allows for a Wick rotation of the propagator to
Euclidean space \cite{leibbrandt1, bassetto}. Although calculations with the
Mandelstam prescription lead to the correct behavior of physical
quantities such as the Wilson line 
\cite{taylor,andrasi,bassetto1}, it is also known 
that this prescription gives rise to some 
unphysical features such as nonlocal ultraviolet divergent terms in
loop diagrams \cite{leibbrandt1}. We would, therefore, like
to study systematically the structure of propagators in the light-cone
gauge using the Mandelstam prescription. This leads to some
interesting and surprising results including the fact that this
prescription induces an unphysical behavior even at the tree level.
More specifically, we find that this prescription
leads to a propagator which has a logarithmic growth for large values of
$\tilde{n}\cdot x$. Following our earlier argument, this would
correspond to the fact that the Mandelstam prescription does not
completely fix the light-cone gauge and we determine this residual
gauge symmetry explicitly. The
second result that comes out of our analysis is that the
light-cone gauge and the Mandelstam prescription induce a coordinate
dependent 
mass for the $\tilde{n}\cdot A$ component of the gauge field which can
be seen from the analytic structure of the retarded propagator as well
as from the
spectral function. This unexpected behavior turns out to be a
consequence of the presence of the residual gauge symmetry in the theory.

Our paper is organized as follows. In section {\bf 2}, we study
the Feynman propagator in the coordinate space with the Mandelstam
prescription. We show that the propagator is not well behaved for
large values of $\tilde{n}\cdot x$. Following our earlier observations
\cite{das}, 
we show that this is a reflection of the fact that the Mandelstam
prescription does not fix the light-cone gauge completely and that the free
theory has a residual local symmetry. In section {\bf 3}, we study in
some detail the structure of the retarded
propagator, which indicates that in the light-cone gauge
with the Mandelstam prescription, a coordinate dependent mass is
induced for the $\tilde{n}\cdot A$ mode. This conclusion is further
supported by the structure of the spectral function. In 
section {\bf 4}, we argue that such an unphysical behavior may be eliminated
through an appropriate gauge fixing of the residual gauge degrees of
freedom of the theory.

\section{Feynman Propagator in the Coordinate Space}

Let us begin with some useful notation for carrying out calculations
in the light-cone gauge. Let $n^{\mu}, \tilde{n}^{\mu}$ represent two
light-like vectors such that $n^{2}=0=\tilde{n}^{2}$, but
$n\cdot\tilde{n}\neq 0$. For simplicity, we will choose sgn $(n\cdot
\tilde{n})$ to be positive. Any vector can now be decomposed as

\begin{equation}
V_{\mu} = \frac{\tilde{n}_{\mu}}{(n\cdot \tilde{n})}\ (n\cdot V) +
\frac{n_{\mu}}{(n\cdot \tilde{n})}\ (\tilde{n}\cdot V) + V_{\mu}^{\rm
  T},\label{decomposition}
\end{equation}
where
\begin{equation}
n\cdot V^{\rm T} = 0 = \tilde{n}\cdot V^{\rm T}.
\end{equation}

Let us next note that Mandelstam's prescription
(\ref{mandelstam}) 
leads upon Fourier transformation to the Green's function
\begin{equation}
 G (x) = \frac{i}{[n\cdot \partial]}\ \delta^{4} (x) = \int
\frac{\mathrm{d}^{4}p}{(2\pi)^{4}}\ \frac{e^{i p\cdot x}}{n\cdot p
  + i \epsilon \tilde{n}\cdot p} = \frac{1}{2\pi}\ \frac{1}{n\cdot
  x - i\epsilon \tilde{n}\cdot x}\ \delta^{2} (x^{\rm T}).\label{FT}
\end{equation}
The explicit representation (\ref{FT}) is very useful in
evaluating systematically the coordinate representation of any
quantity. For example, for a general function $f(p)$, we can write
\begin{equation}
\int \frac{\mathrm{d}^{4}p}{(2\pi)^{4}}\ \frac{f(p)}{n\cdot p +
  i\epsilon \tilde{n}\cdot p}\ e^{ip\cdot x} = \int \mathrm{d}^{4}x'\
  G(x-x') \int \frac{\mathrm{d}^{4}p}{(2\pi)^{4}}\ f(p) e^{ip\cdot
  x'}.
\end{equation} 
In particular, this leads to the following coordinate representation for the
Feynman propagator in the light-cone gauge with the Mandelstam
prescription
\begin{eqnarray}
D_{\mu\nu}^{\rm (PI)} (x) & = &\!\!\! -
\frac{1}{(2\pi)^{2}}\!\left[\!\!\left(\eta_{\mu\nu} 
  - \frac{n_{\mu}\tilde{n}_{\nu}+n_{\nu}\tilde{n}_{\mu}}{n\cdot
    \tilde{n}} + \frac{2(x_{\mu}^{\rm T} n_{\nu} + x_{\nu}^{\rm T}
    n_{\mu}) (\tilde{n}\cdot x)}{(n\cdot \tilde{n}) x^{\rm T}\cdot
    x^{\rm T}} - \frac{2 (\tilde{n}\cdot x)n_{\mu}n_{\nu}}{(n\cdot
    \tilde{n}) (n\cdot x)} \right)\! \frac{1}{x^{2} -
    i\epsilon}\right.\nonumber\\
 &  & \qquad\qquad \left. + \frac{n_{\mu}n_{\nu}}{(n\cdot x)^{2}}\ \ln
  \left(\frac{-x^{2} + i\epsilon}{(x^{\rm
      T})^{2}}\right)\right],\label{feynman}
\end{eqnarray}
where we have identified $(x^{\rm T})^{2} = - x^{\rm T}\cdot x^{\rm T}
\geq 0$. There are several things to note from the explicit structure
of the Feynman propagator (\ref{feynman}). We see that, apart from the first
two  tensor
structures, all other terms vanish for $\tilde{n}\cdot x =
0$. This may be understood by noting that, since (\ref{FT}) is formally
an integration operator, one may evaluate several terms in the Feynman
propagator (\ref{pipropagator}) (with $n^{2}=0$) as
\begin{equation}
\int \frac{\mathrm{d}^{4}p}{(2\pi)^{4}}\ \frac{e^{i p\cdot
  x}}{p^{2} + i\epsilon}\ \frac{1}{n\cdot p + i \epsilon
  \tilde{n}\cdot p} = \frac{i}{n\cdot \tilde{n}} \int_{\tilde{n}\cdot
  x_{0}}^{\tilde{n}\cdot x} \mathrm{d} (\tilde{n}\cdot x') \int
  \frac{\mathrm{d}^{4}p}{(2\pi)^{4}}\ \frac{e^{ip\cdot x'}}{p^{2} +
  i\epsilon},\label{9}
\end{equation}
where the ``prime'' in $x'$ in the exponent refers only to the
coordinate $\tilde{n}\cdot x'$.
We note that if $\tilde{n}\cdot x =0$, the exponent in the first
expression has no dependence
on $(n\cdot p)$ and since the two poles in the integrand of this
expression  lie on the
same side of the complex $(n\cdot p)$ plane, such an integral will
give zero in 
this limit. This implies, from the second expression in
(\ref{9}), that in this case the reference
point can be chosen to be the origin \cite{bassetto2}, which therefore
explains the vanishing of the above terms at $\tilde{n}\cdot x =0$. 

On the other hand, for large values of
$\tilde{n}\cdot x$, the propagator in (\ref{feynman}) has a
logarithmic growth and, therefore, it is
not well behaved. This mildly singular behavior can be viewed from
various points of view. Probably the simplest way is to note that the
prescription (\ref{mandelstam}) does not quite regularize the
singularities when $(n\cdot p), (\tilde{n}\cdot p) \rightarrow 0$
simultaneously. This lack of a bounded behavior of the propagator even with the
Mandelstam prescription is consistent with our analysis and following
our earlier arguments should correspond to the presence of some
residual gauge 
symmetry. To see, in the simplest way, that the theory has a residual
gauge 
symmetry even with the Mandelstam prescription, let us write an
effective Lagrangian density for the free theory which would reproduce the
Mandelstam prescription in the propagator naturally. It is easy to
check that the free Lagrangian density (the non-Abelian theory would
involve a trace over the matrix indices)
\begin{equation}
{\cal L} = - \frac{1}{4} (\partial_{\mu} A_{\nu} - \partial_{\nu}
A_{\mu}) (\partial^{\mu}A^{\nu} - \partial^{\nu} A^{\mu}) -
\frac{1}{2} (N\cdot A)\ \frac{\Box^{2}}{\xi \Box^{2} + \epsilon^{2}
  (\tilde{n}\cdot \partial)^{2}}\ (N\cdot A),\label{effective}
\end{equation}
where
\begin{equation}
N^{\mu} = n^{\mu} + i\epsilon\ \frac{(\tilde{n}\cdot
  \partial)\partial^{\mu}}{\Box},\label{V}
\end{equation}
leads, in the limit $\xi\rightarrow 0$, to the Feynman propagator in
the light-cone gauge with the Mandelstam prescription
\cite{veliev}. The fact that this effective Lagrangian density is not
Hermitian could signal certain difficulties in a theory
incorporating the Mandelstam prescription. However, from the point
of view of looking for 
residual symmetries, the Lagrangian density (\ref{effective}) is suitable
for our purpose. We note that under a gauge
transformation of the form 
\begin{equation}
A_{\mu} (x) \rightarrow A_{\mu} (x) + \partial_{\mu} \omega (x),
\end{equation}
the invariant action will, of course, not change. Furthermore, it
follows from the structure in (\ref{V}) that, under such a
transformation
\begin{equation}
N\cdot A \rightarrow N\cdot A,
\end{equation}
if $\omega (x)$ is a function only of $x^{\mu\, {\rm T}}$. Thus, the
free theory (\ref{effective}) has a residual 
gauge symmetry with a parameter $\omega (x^{\rm T})$. As we have
argued in our earlier work \cite{das}, the presence of this residual
symmetry is responsible
for the propagator (\ref{feynman}) in the light-cone gauge with the
Mandelstam prescription having a (mildly) singular behavior for large
values of $\tilde{n}\cdot x$.

\section{Retarded Propagator and the Spectral Function}

The Feynman propagator (\ref{feynman}) is an analytic function in the
upper half of the complex $\tilde{n}\cdot x$ plane except for an isolated
pole and a logarithmic branch cut beginning at the light-cone
$x^{2}=0$.  Therefore, one can easily write
down a dispersion relation in the complex $\tilde{n}\cdot x$ plane
with one subtraction because of the logarithmic behavior. However, we
do not give any further detail on this and instead we now analyze the
retarded propagator, which is more useful for a better
understanding of the structure of this theory. 

The retarded propagator of the theory can be obtained from the form of
the Feynman propagator in (\ref{feynman}) as \cite{bj,roman}
\begin{eqnarray}
D_{\mu\nu}^{\rm (PI)\ (R)} (x) & = &  2\theta
  (x^{0}) {\rm Im} (D_{\mu\nu}^{\rm (PI)})\nonumber\\
 & = & - \frac{\theta(x^{0})}{2\pi}\left[\left(\eta_{\mu\nu} -
  \frac{n_{\mu}\tilde{n}_{\nu}+ n_{\nu}\tilde{n}_{\mu}}{(n\cdot
  \tilde{n})} + \frac{2(x_{\mu}^{\rm T}n_{\nu} + x_{\nu}^{\rm
  T}n_{\mu}) (\tilde{n}\cdot x)}{(n\cdot \tilde{n}) x^{\rm T}\cdot
  x^{\rm T}}\right)\delta (x^{2})\right.\nonumber\\
 &  & \qquad\qquad \left. - n_{\mu}n_{\nu} \frac{(x^{\rm T})^{2}}{(n\cdot
  x)^{2}}\left(\delta (x^{2}) - \frac{1}{(x^{\rm T})^{2}}\ \theta
  (x^{2})\right)\right].\label{retarded}
\end{eqnarray}
The $\delta (x^{2})$ term simply corresponds to a massless pole in the
momentum space. However, the coefficient of $n_{\mu}n_{\nu}$ shows a
more interesting structure of $\theta (x^{2})$. To appreciate what
this corresponds to, let us define
\begin{equation}
\Delta (x) = \frac{ {\rm sgn} (x^{0})}{2\pi}\ \frac{(n\cdot \tilde{n})
    (x^{\rm
    T})^{2}}{(n\cdot x)^{2}}\left(\delta (x^{2}) - \frac{1}{(x^{\rm
    T})^{2}}\ \theta (x^{2})\right),\label{delta}
\end{equation}
so that we can identify
\begin{equation}
-\frac{\tilde{n}^{\mu}\tilde{n}^{\nu}}{n\cdot \tilde{n}}
D_{\mu\nu}^{\rm (PI)\ (R)} (x) = -\theta
(x^{0})\ \Delta = - \frac{\theta (x^{0})}{2\pi}\ \frac{ (n\cdot
  \tilde{n}) (x^{\rm
    T})^{2}}{(n\cdot x)^{2}}\left(\delta (x^{2}) - \frac{1}{(x^{\rm
    T})^{2}}\ \theta (x^{2})\right).
\end{equation}
Comparing this with the retarded propagator for a massive scalar field
near the light-cone \cite{bj,roman},
\begin{equation}
D^{\rm (R)} (x) = - \frac{\theta (x^{0})}{2\pi}\left(\delta (x^{2}) -
\frac{m}{2\sqrt{x^{2}}}\ J_{1} (m\sqrt{x^{2}})\ \theta (x^{2})\right)
\approx - \frac{\theta (x^{0})}{2\pi}\left(\delta (x^{2}) -
\frac{m^{2}}{4}\ \theta( x^{2})\right),
\end{equation}
we conclude that near the light-cone, the $\tilde{n}\cdot A$ component
of the gauge field, in the light-cone gauge with the Mandelstam
prescription, has developed a coordinate dependent mass
\begin{equation}
m^{2} = \frac{4}{(x^{\rm T})^{2}} > 0.\label{mass}
\end{equation}

We can derive further support for this by looking at the spectral
function associated with the $\tilde{n}\cdot A$ component of the
gauge field. Let us recall that the spectral function \cite{bj,roman} may
be defined as
\begin{equation}
\frac{i}{n\cdot \tilde{n}}\left[\tilde{n}\cdot A (x), \tilde{n}\cdot A
  (y)\right] = \Delta 
(x-y) = \frac{(n\cdot \tilde{n}) (z^{\rm T})^{2}}{(n\cdot z)^{2}}\ 
i\int \frac{\mathrm{d}^{4}p}{(2\pi)^{3}}\ \left(e^{-i p\cdot
  z} - e^{i p\cdot z}\right)\ \rho,\label{commutator}
\end{equation}
where for simplicity of notation we have identified $z^{\mu} =
(x^{\mu}-y^{\mu})$. From the structure of (\ref{delta}),
we note that the covariant commutator (\ref{commutator}) vanishes for
space-like separations, which is consistent with causality. Using the
form of $\Delta$ in (\ref{delta}), we can determine that 
\begin{equation}
\rho = \frac{m\sqrt{z^{2}}}{2J_{1} (m\sqrt{z^{2}})}\ \theta (p_{0})
\delta (p^{2}-m^{2}) \approx \theta (p_{0}) \delta (p^{2} -
m^{2}),\label{spectral} 
\end{equation}
where the second equality holds near the light-cone with $m^{2} =
\frac{4}{(z^{\rm T})^{2}} > 0$. This can be compared
with (\ref{mass}). We see that near the
light-cone, where the discontinuity of $\tilde{n}^{\mu}\tilde{n}^{\nu}
D_{\mu\nu}^{\rm (PI)\ (R)}$ occurs, the spectral
function is positive as it should be. (We emphasize here 
that the spectral function in (\ref{spectral}) is not the inverse
Fourier sine transform of $\left(\delta (z^{2}) - \frac{1}{(z^{\rm
    T})^{2}}\ \theta (z^{2})\right)$; the latter, in fact, is an
ill-behaved function which is ambiguous and regularization dependent.) 

We note that all our results are manifestly invariant under
translations. However, a coordinate dependent mass is quite
unexpected. In fact, the spectral function $\rho$ is normally a
function of the momentum variables whereas in the present case, it
depends on the transverse coordinates (in a translational invariant
manner) as well. The normal assumptions in the derivation of the
spectral representation include invariance of the vacuum under
translations, namely,
\begin{equation}
e^{-iP\cdot x} |0\rangle = |0\rangle,\label{assumption}
\end{equation}
and our result suggests that this assumption in the present case may
be  violated for the $\tilde{n}\cdot A$ modes.

\section{Discussion}

As we have seen, the logarithmic term in the
propagator leads, through its $\theta (x^{2})$ discontinuity, to a
coordinate dependent mass which would necessitate a redefinition of the
vacuum. This behavior is highly unreasonable considering that we are dealing
with a free theory. On the other hand, we have shown that the growth for large
$\tilde{n}\cdot x$ of this logarithm is a
consequence of a residual gauge symmetry (with parameter $\omega
(x^{\rm T})$) present in the theory. From our earlier results
\cite{das}, it follows that if one would fix this residual
symmetry by imposing an appropriate extra gauge condition, this will
lead to a 
propagator which is well behaved at infinity. For example, one could improve
the behavior at infinity by adding to the effective action an extra gauge
fixing term of the form
\begin{equation}
S_{\rm extra} = \lim_{\eta\to 0}\ - \frac{1}{2\eta}\int
\mathrm{d}^{4}x\,\delta(\tilde{n}\cdot x-\tau)(\partial \cdot A)^2 ,
\label{eq22a}
\end{equation}
which is defined at some fixed value of $\tilde{n}\cdot x=\tau$. This
may then remove
from the propagator the logarithmic term together with its $\theta
(x^{2})$ discontinuity We have not worked out the complete expression
for such a gauge-fixed propagator, whose form is rather complicated
and beyond the scope of this brief report. But we expect that the
above behavior would also remove from the theory the unphysical
coordinate dependent mass. One may
understand this feature in a simple way by considering the classical
equations of motion. If we 
decompose the vector potential as in (\ref{decomposition}), it is easy
to check that the Maxwell equation
\begin{equation}
\partial_{\mu} F^{\mu\nu} = \left(\eta^{\mu\nu} \Box -
\partial^{\mu}\partial^{\nu}\right) A_{\nu} = 0,
\end{equation}
leads, in the light-cone gauge, to the component equations
\begin{eqnarray}
\Box \left(\tilde{n}\cdot A\right) & = & \tilde{n}\cdot \partial
\left(\partial\cdot A\right),\nonumber\\
\Box A_{\mu}^{\rm T} & = & \partial_{\mu}^{\rm T} \left(\partial\cdot
A\right). 
\end{eqnarray}
At first sight, it would seem that these do not yield massless
equations  for the components of the gauge field. However, by
choosing an appropriate gauge parameter $\omega (n\cdot x,
x^{\rm T})$ such that $\partial\cdot A = 0$  at some fixed
$\tilde{n}\cdot x$, the extra terms in the above equations can be
eliminated. 

In summary, we have shown that, at the tree level, the propagator in
the light-cone  gauge
with the Mandelstam prescription has a logarithmic growth for large
values of $\tilde{n}\cdot x$ which is a consequence of a residual
gauge symmetry. In turn, this induces an unphysical coordinate dependent
mass for the $\tilde{n}\cdot A$ component of the gauge field. In order to
remove this feature from Green's functions, one must
eliminate the unphysical degrees of freedom by fixing consistently the
residual gauge symmetry, although in the calculation of physical 
$S$-matrix elements, the Mandelstam-Leibbrandt prescription 
seems to be sufficient to obtain the correct behavior of physical
quantities.

\vskip .7cm

\noindent{\bf Acknowledgment:}
\medskip

We would like to thank Prof. J. C. Taylor for  helpful discussions.
This work
was supported in part by the US DOE Grant number DE-FG 02-91ER40685
and by CNPq as well as by FAPESP, Brazil.

\end{document}